\documentclass[eqsecnum,aps,prd,nofootinbib]{revtex4}

\usepackage{graphicx}%
\usepackage{enumerate}%
\usepackage{amsmath}%
\usepackage{amsfonts}
\def\be{\begin{equation}}
\def\bea{\begin{eqnarray}}
\def\ee{\end{equation}}
\def\eea{\end{eqnarray}}

\def\l: { \!:\!\!   }
\def\r: {  \!\!:\!  }
\def\lt: { :\!\!   }
\def\rt: {  \!\!:  }

\def\openone{\leavevmode\hbox{\small1\kern-3.3pt\normalsize1}}%
\def\simle{%  ``less than about'' symbol
    \mathrel{\rlap{\raise 0.511ex
        \hbox{$<$}}{\lower 0.511ex \hbox{$\sim$}}}}

\begin{document}

\title{Unitarity and Holography in Gravitational Physics}

\author{Donald Marolf}

\affiliation{Physics Department, UCSB, Santa Barbara,  \\
CA 93106, USA \\\texttt{marolf@physics.ucsb.edu}}

\begin{abstract}
Because the gravitational Hamiltonian is a pure boundary
term on-shell, asymptotic gravitational fields store information in
a manner not possible in local field theories. This fact has
consequences for both perturbative and non-perturbative quantum
gravity. In perturbation theory about an asymptotically flat
collapsing black hole, the algebra generated by asymptotic fields on
future null infinity within any neighborhood of spacelike infinity
contains a complete set of observables.   Assuming that the same algebra remains complete at the non-perturbative quantum level, we argue that either 1) the
S-matrix is unitary or 2) the dynamics in the region near timelike,
null, and spacelike infinity is not described by perturbative
quantum gravity about flat space. We also consider perturbation
theory about a collapsing asymptotically anti-de Sitter (AdS) black
hole, where we show that the algebra of boundary observables within
any neighborhood of any boundary Cauchy surface is similarly
complete. Whether or not this algebra continues to be complete
non-perturbatively, the assumption that the Hamiltonian remains a
boundary term implies that information available at the AdS boundary
at any one time $t_1$ remains present at this boundary at any other
time $t_2$.
\end{abstract}

\maketitle

\tableofcontents

\section{Introduction}

Arguments for information loss in black hole evaporation are
typically based on locality and causality in quantum field theory on
a fixed background (see e.g. \cite{Hawking1,Wald}). In perturbative
quantum gravity these properties also hold at zeroeth order in the
Planck Length $\ell_p$, where back-reaction is ignored. However, strict
locality explicitly fails at the first interacting order.  A clean
signal of this failure is the fact that a form of time evolution is generated by a
boundary term at spacelike infinity (e.g., the ADM energy in
asymptotically flat space \cite{ADM}).  This feature is closely related to
the lack of local observables in diffeomorphism-invariant theories.

We show below that this simple observation leads to interesting
results. For example, consider the context of perturbation
theory about asymptotically flat collapsing black
hole backgrounds.  There we will show that, at first interacting order 
and beyond, a complete set of observables is contained in
the algebra generated by fields on future null infinity ($I^+$) within any
neighborhood of spacelike infinity ($i^0$).
In the asymptotically anti-de Sitter (AdS) context,
the algebra of boundary observables defined by any neighborhood of a
boundary Cauchy surface is similarly complete in perturbation theory.  As a result, in both cases full information about the quantum state is contained in the
asymptotic fields.

We refer to the above completeness results as `perturbative holography.'  However, we caution the reader that, in contrast to \cite{Lenny}, our use of this term
does not directly imply any particular limit on the number of
degrees of freedom.  The centrality of energy conservation to any
discussion of unitarity was previously emphasized in \cite{BSP},
while the representation of gravitational energy as a boundary term
and the associated ability of the long-range gravitational fields to
store information was emphasized in \cite{BMR}.  The arguments below
stem from a fusion of these ideas.  Other works connecting
energy conservation to black hole unitarity include \cite{MP}.

Before beginning the main arguments, it is appropriate to briefly address three common objections that the reader may already hold:

%\vspace{.3cm}

%\begin{enumerate}[{\bf Object\i on 1:}]

\begin{description}

\item [{\bf Objection \#1,   Locality via gauge fixing:}]
The reader may object that perturbative quantum gravity appears both
local and causal in, say, de Donder gauge.  However, it is important to recall that such gauges contain propagating  longitudinal gravitons associated with residual gauge symmetries.  As is
familiar from the Coulomb gauge in Maxwell theory, gauge fixing all
residual symmetries removes the apparently manifest locality so that no immediate conclusions can be drawn regarding the nature of obsevables. In
Yang-Mills theory one can avoid these issues by constructing Wilson
loops which provide a complete set of compactly supported
observables.   In contrast, no such compactly supported observables are available
in diffeomorphism-invariant theories of gravity.

\item [{\bf Objection \#2,  The characteristic initial value problem:}]  Section \ref{pert} considers perturbations about an asymptotically flat collapsing black hole spacetime. At the level of rigor used below, the characteristic initial value theorem states that the radiative parts of metric perturbations on the future horizon ($H^+$) and future null infinity ($I^+$) form a complete set of independent operators.  As a result, the radiative parts of metric perturbations on $I^+$ cannot, by themselves, define a complete set of observables in this context.  It is important to note that this statement does not contradict our claims.  Indeed, our argument below makes explicit use of {\it both} the radiative and the non-radiative (Coulomb) parts of the metric perturbations on $I^+$. These Coulomb parts are {\it not} independent of the radiative parts of metric perturbations on $H^+$, but are instead related to the full set of radiative perturbations by the gravitational constraints.

\item [{\bf Objection \#3,  Comparison with classical physics:}]  While we use a quantum-mechanical language below, replacing certain commutators by Poisson Brackets suffices to recast our perturbative arguments in the language of classical gravitational physics.    As a result, our arguments imply that in {\it classical} perturbative gravity the Poisson algebra\footnote{In fact, one requires a certain closure of the usual Poisson algebra which allows one to flow any element $A$ of the algebra by a finite amount along the Hamiltonian vector field defined by any other element $B$.} generated  by fields on future null infinity ($I^+$) in any neighborhood of spacelike infinity ($i^0$) contains a complete set of observables, and that a similar result holds in the anti-de Sitter context.

The reader may feel that this statement should contradict the fact the black holes lose information in classical gravity.  That no such contradiction arises can be illustrated using the SO(3) angular momentum generators $J_x,J_y,J_z$.  Of course, $J_z$ lies in the Poisson algebra generated by $J_x, J_y$.  Nevertheless, at the classical level, the ability to measure $J_x$ and $J_y$ imparts no knowledge of $J_z$. Full information is obtained only about {\it algebraic} functions $f(J_x, J_y)$.  It is only at the quantum level that the situation changes, and that alternating measurements of $J_x$ and $J_y$ can indeed provide information about $J_z$.     This last point will be emphasized in a companion paper \cite{soon}, which also resolves a number of possible paradoxes that the reader may fear might be associated with such measurements.

\end{description}
%\end{enumerate}

%\vspace{.3cm}

Having dispensed with the above objections, we may now turn to the main arguments.
At the quantum level our discussion is somewhat formal.  However,  at least in the perturbative context, mathematically rigorous results can be obtained by reinterpreting the arguments below in terms of classical gravity,  replacing commutators with Poisson Brackets and (where appropriate) with finite flows along Hamiltonian vector fields.  As briefly discussed under objection \#3 above, while such results have minimal implications for classical physics, it is clear that they set the stage for more interesting effects at the quantum level.

We begin with the asymptotically flat context in section
\ref{AsFlat}. After deriving perturbative completeness of the algebra near $i^0$, we
consider implications for the non-perturbative theory.  Assuming that the same algebra remains complete in the non-perturbative quantum theory, we show that either 1) the
S-matrix is unitary or 2) the dynamics in the region near timelike,
null, and spacelike infinity is not described by perturbative
quantum gravity about flat space.

We then derive perturbative
holography for asymptotically anti-de Sitter (AdS) quantum gravity in section \ref{AdS}.  We also note that, whether or not the stated algebra continues to be complete non-perturbatively, the
assumption that the Hamiltonian remains a boundary term implies a
form of boundary unitarity.  In particular, information available at
the AdS boundary at any one time $t_1$ remains present at this
boundary at any other time $t_2$. We close with some final
discussion in section \ref{disc}.

\section{Quantum Gravity in Asymptotically Flat Space}
\label{AsFlat}

To avoid making detailed assumptions about the quantum nature of
gravity,  it natural to proceed using either semi-classical methods
or perturbation theory. We choose the latter here, where we have in
mind treating perturbative gravity as an effective field theory (in
which appropriate new parameters may need to be added at each
order). This is the setting for section \ref{pert}.  Section
\ref{nonpert} then studies the implications for the
non-perturbative theory and discusses unitarity of the S-matrix.

\subsection{The Holographic nature of perturbative gravity}
\label{pert}

It is useful to being with a brief summary of the argument:  We
consider perturbation theory around an asymptotically flat classical
solution which is flat in the distant past but contains a black hole
in the distant future.  The argument below simply uses  the Hamiltonian (an operator at $i^0$) to translate any operator on past null infinity ($I^-$) into the distant past, deep into the flat region before the black hole forms.   The perturbative equations of motion then express any such operator in terms of operators on $I^+$. I.e., since the black hole does not form until much later, very little of the operator falls into the black hole.  Furthermore, since we translated the operator on $I^-$ into the distant past, the support on $I^+$ is concentrated near $i^0$.  Taking a limit yields the desired result.

\begin{figure}
\includegraphics[width=5cm] {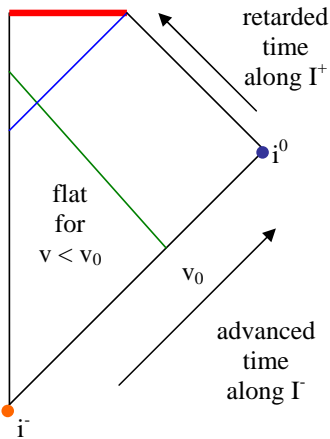}
 \caption{The spacetime is flat before advanced time $v_0$, but the formation of a black hole prohibits a
 regular $i^+$.}
 \label{collapse}
 \end{figure}

It is convenient to perturb about a background solution which is exactly flat
space before some advanced time $v_0$ (see figure \ref{collapse}).  For familiarity and concreteness, we consider pure Einstein-Hilbert
gravity in 3+1 dimensions so that the black hole forms from
gravitational waves arriving from past null infinity ($I^-$).
 Adding matter fields or changing the number of dimensions would not
significantly change the analysis.\footnote{Except that the infrared behavior improves in higher dimensions.  In 3+1 dimensions, our
argument is rather formal in that it ignores infrared divergences
associated with soft gravitons. While it may be interesting to
examine the detailed effect of IR divergences on the argument below,
here we simply assume that the usual techniques \cite{Weinberg}
allow us to use gravitational perturbation theory and to speak of an
S-matrix.  In higher dimensions, no such divergences arise.} The
essential inputs are only diffeomorphism-invariance (so that the Hamiltonian is indeed a boundary term) and our choice
of boundary conditions.

To begin the main argument, let  $\tilde g_{ab}$ denote the metric of the background spacetime and write the
dynamical metric as $g_{ab} = \tilde g_{ab} + \kappa h_{ab}$ where
$\kappa^2 = 8 \pi G$ so that the action for $h_{ab}$ has canonical
kinetic term.
As usual, we work to some finite order in $\kappa$ and
discard terms of higher order.  We will not need to be explicit
about the details below; all that is important is that we work to
some order in which interactions are relevant so that the
gravitational version of Gauss' law leads to a non-trivial
gravitational flux (see \eqref{Phidef} below) at spacelike infinity ($i^0$).
For later use it will also be convenient to expand the
background about flat space by writing $\tilde g_{ab} = \eta_{ab} +
\kappa \tilde h_{ab}$.  The latter expansion is useful near infinity
where $\tilde h_{ab}$ is small.

The perturbations $h_{ab}$ may be quantized in any gauge for which all
propagating modes are physical; e.g. a Coulomb-like gauge. The
Hamiltonian in such gauges is necessarily non-local, but this will
not be a complication.  The advantage of such gauges is that all
equations of motion hold at the level of the Heisenberg operators. For example, the gravitational equivalent of Gauss' law holds as an
operator identity and need not be imposed as a constraint on
physical states.

We now remind the reader of several facts from classical general
relativity.  First, recall that the total energy of the full metric
$g_{ab}$ is given by the Arnowitt-Deser-Misner (ADM) boundary term
at spatial infinity ($i^0$).  We denote this boundary term $\Phi$ as
it will be convenient to think of this term as a gravitational flux.
We have
 \be
 \label{Phidef}
 \Phi = \frac{1}{2\kappa} \int_C dA \  \left( r^a P^{bc} D_b
 - r^b P^{ac} D_b  \right) (\tilde h_{ac} + h_{ac}),
 \ee
where $r^a$ is a radial unit normal, $C$ is a cut of $i^0$ as
defined e.g. in \cite{i0}, $dA$ is the area element on $C$, and
$D_a$ is the covariant derivative defined by the fixed flat metric
$\eta^{ab}$ which also defines the spatial projection $P_{ab}$
orthogonal to the chosen time direction.

Second, if past timelike infinity $(i^-)$ is regular, then the ADM
energy can also be expressed as the integral over past null infinity
($I^-$) of the flux of stress-energy through $I^-$ due to
gravitational radiation (see e.g. \cite{ABR}). This flux is given by
the news tensor, but may be equally well thought of as the integral
of the appropriate component of the stress tensor of linearized
gravity integrated along $I^-$ (see e.g. \cite{CZ}).  Either
expression is purely quadratic in $g_{ab} - \eta_{ab}$, where
$\eta_{ab}$ is a fixed flat metric at infinity. This calculation
shows explicitly that $\Phi$ agrees near $I^-$ with the Hamiltonian of
linearized gravity about flat space, where the linearized field is $\tilde h_{ab} + h_{ab}$. We denote this
Hamiltonian $H^{lin}_{\tilde h + h}$.   Note that since the
perturbations $\tilde h_{ab}, h_{ab}$ fall off at $I^-$, this
linearized Hamiltonian also generates translations along $I^-$ in
the full theory (and in particular at any order in perturbation
theory).

Since $H^{lin}_{\tilde h + h}$ is quadratic, it is straightforward
to expand in powers of $h_{ab}$:
 \be
 \label{expandH}
 H^{lin}_{\tilde h + h} = \tilde E + S +  H^{lin}_{h}.
 \ee
Here $\tilde E$ is the $v$-dependent energy of the background metric $\tilde
g_{ab}$, $S$ denotes a set of ``source terms'' linear in both
$\tilde h_{ab}$ and $h_{ab}$, and $H^{lin}_{h}$ is just the integral
of the (quadratic) stress tensor for perturbations $h_{ab}$
propagating on the flat metric $\eta_{ab}$.

Most importantly for our purposes, the above results can be derived
using the equations of motion near $I^\pm$ expanded only to second
order in $h_{ab}$.  As a result, they hold in perturbative classical
gravity at any order beyond the free linear theory; i.e., at any
order where the gravitational Gauss' law makes $\Phi$ non-trivial.
Furthermore, the results also hold in perturbative quantum gravity
as the only operator that requires regularization is the (quadratic)
stress tensor for gravitons propagating in flat space.

Below, it will be convenient to denote operators on $I^-$ as
$h_{ab}(v)$, and to speak as if they are well-defined operators. In
doing so we choose a notation which suppresses several details.
First, some rescaling with $r$ is required to define finite objects
on $I^-$.  Second, we implicitly assume that the operators have been
smeared with appropriate test functions. Third, at certain points
below it will be convenient to assume that an expansion in spherical
harmonics has been performed and that each $h_{ab}(v)$ has a
definite angular momentum.

Since we consider perturbations about a background $\tilde g_{ab}$
which is flat before the advanced time $v_0$, past timelike infinity
is regular.  As a result, the relation
 \be
 \label{PAsF}
 \Phi = H^{lin}_{\tilde h + h}
 \ee
holds as an equality of Heisenberg-picture quantum operators.
This relation is somewhat subtle, however, since $\Phi$ as defined in
\eqref{Phidef} is linear in $\tilde h_{ab} + h_{ab}$ while $H^{lin}$ is quadratic.  The point here is simply that $H^{lin}$ is defined by linearizing about a certain background ($\eta_{ab}$).  As a result, the relationship between $\Phi$ and $H^{lin}$ is sensitive to this choice of background.  In particular, subtracting the ($v$-dependent) energy $\tilde E (v)$ of the background metric yields $\Phi - \tilde E(v)  = S(v)  + H^{lin}_h$, where $S(v)$ is an operator linear in both $\tilde h_{ab}(v)$ and $h_{ab}(v)$ as in \eqref{expandH}.  Thus, $S(v)$ has an explicit $v$-dependence though the background $\tilde h_{ab}(v)$.  In contrast, the operator $H^{lin}_h$ is just what would appear in linearized gravity about flat space; $H^{lin}_h$ has no explicit $v$-dependence.

From (\ref{PAsF}) we see that $\Phi$ generates $v$-translations of
$\tilde h_{ab} + h_{ab}$ in the sense that
  \bea
  \label{vtrans}
 (\tilde h_{ab} + h_{ab})(v) &=&  e^{-i \tau \Phi} (\tilde h_{ab}
 + h_{ab})(v- \tau) e^{i \tau \Phi}, \ \ \ \ \  \ \ {\rm or}
  \cr h_{ab}(v) &=&
e^{-i \tau \Phi} h_{ab}(v-\tau) e^{i \tau \Phi} + \tilde h_{ab}
(v-\tau) - \tilde h_{ab}(v).
  \eea
  The terms involving $\tilde h_{ab}$ on the final right-hand-side
are associated with the source terms $S(v)$ in (\ref{expandH}), or
equivalently with the difference between $\Phi$ and $H^{lin}_h$. The role of these c-number terms is to compensate for the fact that $\Phi$ effectively translates both the perturbation and the background.
Equation (\ref{vtrans}) is a key result which we will use liberally.
Note that while $\tilde h_{ab}(v)$ is formally of order $1/\kappa$,
its effects become arbitrarily small at sufficiently large $r$; i.e., near
infinity terms involving $\tilde h_{ab}(v)$ need not interfere with
our perturbative treatment.

We now proceed to our main argument.  Choose any retarded time $u_1$
along $I^+$ and any operator $h_{ab}(v)$ at any advanced time $v$ on
$I^-$.  We wish to show that, in any state, the operator $h_{ab}(v)$
can be arbitrarily well approximated by elements of the algebra
${\cal A}^+_{u_1}$ generated by operators at $I^+$ supported at
retarded times $u < u_1$.  By convention\footnote{We could also have
used the Bondi energy associated with a cut of $I^+$ at retarded time $u$
to approximate $\Phi$ as $u \rightarrow - \infty$, but our argument
loses nothing by making the above simplifying convention.}, we
consider $i^0$ to be a point on $I^+$ with $u = - \infty$ so that
${\cal A}^+_{u_1}$ contains $\Phi$.  Since we may use
(\ref{vtrans}), it remains only to approximate $h_{ab}(v-\tau)$ by
operators in ${\cal A}^+_{u_1}$.

To do so, note that since $\tilde g_{ab}$ is flat for $v < v_0$,
there is some advanced time $v_1(u_1,L)$ such that all null
geodesics (with angular momentum $L$) launched from $I^-$ before
$v_1(u_1,L)$ arrive at $I^+$ before retarded time $u_1$.  As a
result, in the geometric optics approximation to the linearized
theory, the equations of motion relate operators $h_{ab}(v-\tau)$
with angular momentum $L$ and $v - \tau < v_1(u_1,L)$ to an operator
in ${\cal A}^+_{u_1}$. This situation is summarized in figure 2.

Beyond the geometric optics approximation,
and taking into account non-linear corrections at some fixed order
of perturbation theory, we may use the equations of motion to write
 \be
  h_{ab}(v-\tau) =  {\cal O}_{ab}(v-\tau, u_1) + \Delta_{ab}(v-\tau,
  u_1),
   \ee
where ${\cal O}_{ab}(v-\tau, u_1) \in {\cal A}^+_{u_1}$ and
$\Delta_{ab}(v-\tau,u_1)$ is an error term.  Because all corrections
are determined by Green's functions peaked on the light cone, in any
fixed state (having a finite number of particles on $I^-$) the error
$\Delta_{ab}(v-\tau,u_1)$ will vanish as some power law in the limit $v_1 - (v - \tau) \rightarrow \infty$.

\begin{figure}
 \includegraphics[width=5cm] {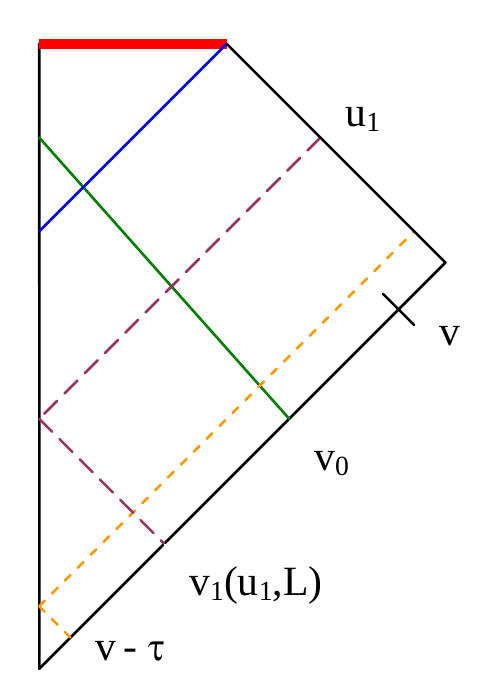}
 \caption{As $\tau \rightarrow +\infty$, operators at $v-\tau$ on $I^-$
 can be written in terms of operators on $I^+$ before retarded time $u_1$.}
 \label{translate}
 \end{figure}

This is nearly the desired result.  For the final step of the argument, it is useful to express $\Delta_{ab}(v-\tau,u_1)$ in terms of the operators $h_{ab}(v)$ on $I^-$ using the same perturbative equations of motion.
The largest contributions will come from the region near $v_1(u_1,L)$, but there will be power-law suppressed contributions from other regions as well.  Now, since we observed above that matrix elements of $\Delta_{ab}(v-\tau,u_1)$ must vanish as a power law in all states having a finite number of particles on $I^-$ in the limit $(v-\tau) \rightarrow \infty$, we may expand $\Delta_{ab}(v-\tau,u_1)$ in powers of $(v-\tau)^{-1}$; i.e., we write

\begin{equation}
\label{expD}
\Delta_{ab}(v-\tau,u_1) \approx \sum_{n >0} (v-\tau,u_1)^{-n} \Delta^{(n)}_{ab}(u_1),
\end{equation}
where the operators $\Delta^{(n)}_{ab}(u_1)$ are independent of $v, \tau$.  We will use \eqref{expD} only as an asymptotic series and do not require convergence.  As is described in more detail in appendix \ref{app}, at any fixed order in perturbation theory, the operators $\Delta^{(n)}_{ab}(u_1)$ are simply integrals over products of operators $h_{ab}$ on $I^-$ with a weighting function determined by $u_1$.  Using (\ref{expD}), consider now the contribution
\begin{equation}
\label{transD}
e^{-i \tau \Phi} \Delta_{ab}(v-\tau,u_1)e^{i \tau \Phi} =  \sum_{n > 0} (v-\tau,u_1)^{-n} e^{-i \tau \Phi} \Delta^{(n)}_{ab}(u_1) e^{i \tau \Phi}
\end{equation} of $\Delta_{ab}(v-\tau,u_1)$ to \eqref{vtrans}.  We wish to take the limit $\tau \rightarrow \infty$.  This has two effects.  First, the factors of
$e^{\pm i  \tau \Phi}$ translate each $\Delta^{(n)}_{ab}(u_1)$ toward $i^0$.  Since
correlation functions in any Fock space state approach those of the
vacuum at large times, the large $\tau$ limit of each $e^{-i \tau \Phi} \Delta^{(n)}_{ab}(u_1) e^{i \tau \Phi}$ is a (finite) c-number determined by the background metric $\tilde g_{ab}$.  It follows that the large $\tau$ limit of \eqref{transD} must vanish due to the factors of $(v-\tau,u_1)^{-n}.$

Combining the above results we have
 \be
  h_{ab}(v) = \lim_{\tau \rightarrow \infty} \bigl[ e^{-i \tau \Phi}{\cal O}_{ab}(v-\tau, u_1) e^{i \tau
\Phi} + \tilde h_{ab}(v - \tau) - \tilde h_{ab}(v) \bigr]. \ \ \ \ \ \ \
 \ee
Since any c-number (e.g., $\tilde h_{ab}(v)$ or $\tilde h_{ab}(v)$)  lies in ${\cal A}^+_{u_1}$, the right-hand side
contains only elements of ${\cal A}^+_{u_1}$ as desired.  Thus we have
shown that any fundamental field on $I^-$ can be expressed with
arbitrary accuracy as an element of ${\cal A}^+_{u_1}$.  Similarly,
any product of such fields can be expressed (with arbitrary
accuracy) by taking the above limit separately for each operator in
the product.

We conclude that a complete set of operators on $I^-$ is contained
in the weak closure of ${\cal A}^+_{u_1}$. For convenience, we used
a Coulomb-like gauge, but the corresponding result for
gauge-invariant observables follows immediately in any gauge.

\subsection{Non-perturbative gravity and Unitarity of the S-matrix}
\label{nonpert}

We saw above that perturbative gravity about an asymptotically flat spacetime
is holographic in the sense that the algebra of observables generated by the ADM Hamiltonian $\Phi$ and the usual asymptotic fields within any neighborhood of $i^0$ in $I^+$ contains a
complete set of observables. Thus, all of the information present at
$I^-$ is encoded in observables in the stated region of $I^+$.   However,
discussions of black hole unitarity typically focus on unitarity of
the S-matrix. This is a somewhat different question, defined in
terms of the Fock spaces at $I^\pm.$ In particular, it is manifestly
clear that, at a finite order in perturbation theory about a
collapsing black hole, the Fock spaces at $I^\pm$ do not encode the
same degrees of freedom.

From our point of view, this difference arises because there is no
regular future timelike infinity in a black hole spacetime.  As a
result, in perturbation theory about such a background, the total
gravitational flux $\Phi$ cannot be expressed solely in terms of the
stress tensor at $I^+$, and thus cannot be expressed in terms of
creation and annihilation operators at $I^+$.  This was possible at
$I^-$ only due to the particular boundary conditions chosen at
$i^-$.

On the other hand, one expects any black hole
that forms to decay by Hawking evaporation.  While this process cannot be fully described in perturbation theory,  perturbative quantum gravity (say, about flat spacetime) may well be a good description of the end products resulting from the decay.  In this case,  $i^+$ is regular. Let
us therefore suppose that, in any asymptotically-flat state of the non-perturbative
theory, perturbative quantum gravity about flat spacetime becomes an
arbitrarily good approximation for field operators near
past ($i^-$ and $I^-$), future ($i^+$ and $I^+$), and spacelike
infinity ($i^0$).   Let us also extrapolate our perturbative result and assume that the algebra generated by $\Phi$ and asymptotic fields on $I^+$ in any ${\cal A}^+_{u_1}$ again contains a complete set of observables, at least within an appropriate superselection sector\footnote{Note that this is necessarily a new assumption.  In particular, it does not follow from the assumption that perturbation theory is arbitrarily good near infinity.  Our previous perturbative argument required us to propagate fields from $I^-$ to $I^+$ through the bulk of the spacetime where non-perturbative effects can be important.  The purpose of mentioning our perturbative argument here is only to render this assumption plausible by removing objections based on perturbative fields falling into semi-classical black holes.  See e.g. \cite{BadH} for further discussion of the idea that this assumption may hold only within an appropriate superselection sector.}. Since we have a regular $i^+$, the gravitational flux  $\Phi$ {\em can} be
expressed as the integral of the linearized stress tensor over
$I^+$. It follows that any observable can indeed be expressed in
terms of creation and annihilation operators on $I^+$.  Our discussion is tailored to settings with no stable massive particles but, since we assume that physics is perturbative near $i^+$, allowing stable massive particles would merely require $\Phi$ to be expressed in terms of the stress tensor at both $I^+$ and $i^+$, and for the corresponding creation and annihilation operators at $i^+$ to be included in our discussion.

Note that the other Poincar\'e generators on $I^-$ can be related to
those on $I^+$ in precisely the same manner as was done for
time-translations.  Thus the Poincar\'e-invariant vacuum on $I^-$
also defines a Poincar\'e-invariant state on $I^+$.  Since such a
state is unique in perturbative quantum field theory,
the Fock vacua on $I^\pm$ coincide.

The unitarity of the S-matrix now follows in the usual way.
$N$-particle states are defined by the action of local operators at
$I^\pm$ on the Fock vacuum.  Since local operators can be translated between $I^+$
and $I^-$, and since the vacuua at $I^\pm$ coincide, these constructions merely define two bases for the same
Hilbert space.  The $S$-matrix is then nothing more than the expression of the dictionary between $I^-$ and $I^+$. Since the two bases define the same Hilbert space, the S-matrix is unitary.

\section{Asymptotically AdS Quantum Gravity}
\label{AdS}

We saw above that there is a sense in which perturbative gravity is
holographic in asymptotically flat space.  As we now show, similar
methods lead to an analogous result in the context of (e.g., 3+1)
AdS asymptotics.  To be specific, we require that the metric has a
Fefferman-Graham expansion \cite{FG} (see also \cite{HS}) of the
form
 \be
 g_{ab} = \frac{\ell^2}{r^2}dr^2 + \left( g_{(0)ij} \frac{r^2}{\ell^2} + g_{(1)ij} \frac{r}{\ell} + g_{(2)ij} + g_{(3)ij} \frac{\ell}{r}
  + \dots \right) dx^i dx^j,
 \ee
for some fixed boundary metric $g_{(0)ij}$. Here $\ell$ is the AdS
scale, the $x^i$ are coordinates on $S^2 \times {\mathbb R}$, and
the $\dots$ represent higher order terms in $r/\ell$ which may
include cross terms of the form $dr dx^i$.
 The coefficients
$g_{(1)ij},g_{(2)ij}$ are determined by the choice of $g_{(0)ij}$
(and any matter fields, see below) via the Einstein equations. In
contrast, $g_{(3)ij}$ depends on the propagating degrees of freedom
in the bulk. For convenience below, we will take one of the coordinates to be some $t$ such that the intersection of each $t=constant$ surface with the boundary spacetime is a Cauchy surface of the boundary spacetime.

Certain simplifications arise if we couple the gravitational field
to a conformally coupled scalar field $\phi$, though this does not
appear to be essential to the argument. In 3+1 dimensions we take
the scalar to have the standard asymptotic behavior (see e.g.
\cite{BF})
 \be
 \phi = \frac{\alpha}{r} + \frac{\beta}{r^2} + \dots ,
 \ee
where $\alpha$ will be a fixed scalar function on the boundary.   In
this context, we may fix $g_{(0)ab}$ to be the metric on the
Einstein static universe.  We also take $\alpha =0$
 before some time $t_i$ and again after some
time $t_f$.  In particular, we take the background metric $\tilde
g_{ab}$ to describe empty AdS space to the past of some boundary
time $t_f$. For $t_i < t < t_f$, the time-dependence of $\alpha$
will be chosen to generate scalar radiation which collapses to form
a black hole\footnote{It is straightforward to find such boundary
conditions.  Consider for the moment a solution to the free
conformally-coupled scalar wave equation on the 3+1 Einstein static
universe in which $\phi =0$ in the northern hemisphere at some time
$t_i$, but in which a large spherically-symmetric pulse of
short-wavelength scalar radiation crosses the equator a short time
later. Now restrict this solution to the northern hemisphere and
conformally map the result to a solution of the free scalar equation
on AdS. The $\alpha(x)$ defined by this solution generates a large
spherical pulse of scalar radiation which enters the AdS space
through the boundary shortly after time $t_i$.  For large enough
amplitude, this pulse will collapse to form a black hole.}.  Note
that for such boundary conditions we may define a time-dependent
Hamiltonian which differs from the Hamiltonian for $\alpha =0$ by
the addition of certain source terms for the scalar field in the
region $t_i < t < t_f$.

Now consider any spacelike surface $\Sigma$ in the initial pure AdS
region. It is clear that any field at any later time can be
expressed in terms of fields on $\Sigma$.  Similarly, {\em in the
linearized approximation}, any field on $\Sigma$ can be expressed in
terms of the boundary fields $g_{(3)ab}$ and $\beta$ at earlier
times. Some explicit formulae for the scalar case\footnote{The
explicit formula in \cite{explicit} express local bulk fields in
terms of boundary fields in a compact region of of the boundary
causally disconnected from the point at which the local bulk field
is defined.  A small additional time translation will reexpress this
result in terms of fields at earlier times.} appear in e.g.
\cite{explicit}, but the fact that this is possible follows
immediately from the observation that any linearized solution with
given $\delta g_{0(ab)}$ and $\alpha$ is determined by the values of
$\delta g_{0(ab)}$, $\alpha$, $g_{3(ab)}$, and $\beta$ to the past
of $\Sigma$. This in turn follows from a simple argument: Suppose
that two such solutions have the the same values of $\delta
g_{0(ab)}$, $\alpha$,$g_{3(ab)}$, and $\beta$ to the past of
$\Sigma$, so that their difference has $\delta g_{0(ab)} = \alpha =
g_{3(ab)} = \beta =0.$ This solution also satisfies ingoing boundary
conditions, and so must vanish in the distant past. In particular,
the energy function defined by Dirichlet boundary conditions vanishes in the distant past when evaluated on this solution.  But by construction our
difference solution conserves this notion energy, so that it must vanish at all
times; i.e., the solution must vanish identically.  We conclude that any linearized field on $\Sigma$ is determined by the boundary fields $g_{(3)ab}$ and $\beta$
at earlier times.  As a result, any operator in the linearized theory may be expressed in terms of the boundary operators $g_{(3)ab}$ and $\beta$.   

It follows that the same result holds at each order in perturbation theory.
However, we stress that since we have used $g_{(3)ab}$ and $\beta$ at all times, this statement does yet not constitute ``holography."  Instead, it merely notes certain properties of wave equations in anti-de Sitter space\footnote{We thank Stefan Hollands for pointing out that this result is similar to certain consequences of Holmgren's uniqueness theorem \cite{Hor}, though in our context we find global uniqueness of the solution.}.

To complete the argument for our perturbative holography, simply note that the algebra of
boundary operators ${\cal A}_{t, \Delta t}$ supported within any
time $\Delta t$ of any boundary time $t$ contains the Hamiltonian.
Thus we may in fact express any perturbative field on $\Sigma$ as an
element of ${\cal A}_{t, \Delta t}$ for {\em any} $t, \Delta t$, including
those times in the distant future. For $t
> t_i$, we need merely include the effects of the source terms in
the time-dependent Hamiltonian.  Since the coordinate $t$ is arbitrary, it follows that the algebra
generated by boundary fields within any neighborhood of any boundary
Cauchy surface is similarly complete.

At least at the level of perturbation theory, we have expressed any
observable in terms of the boundary fields at an arbitrary time $t$.  In this
sense, perturbative gravity in AdS may be called ``holographic."
However, as in the case of asymptotically flat space, this
observation does not immediately allow us to express our observable
as a set of standard creation and annihilation operators at the
desired late time.  As in flat space, it is manifestly clear that such an expression is {\it not} possible at any finite order in perturbation theory about a black hole
background.

Let us therefore briefly consider a non-perturbative theory.  In
asymptotically flat space we assumed that perturbative quantum
gravity was a good approximation at both early and late times in order to derive unitarity of the S-matrix.  We could give a similar argument in the AdS case, but it would require non-standard boundary conditions that allow the particles to leave the original AdS space.   E.g., we could consider the evaporon model of \cite{Jorge}.  However, it is perhaps more enlightening to maintain standard AdS boundary conditions and to derive a more restrictive result.  To proceed, we assume only that

\begin{enumerate}[i)]

\item  There is a well-defined, perhaps time-dependent, family of
self-adjoint operators $H(t)$.

\item  Each $H(t)$ is a member of the corresponding algebra
${\cal A}_{t, \Delta t}$ of boundary observables.

\item  This family of operators generates time evolution in
the usual sense associated with time-dependent Hamiltonians; i.e.,
the time translation is $U(t_1,t_2) = {\cal P}\exp \left( -i
\int_{t_1}^{t_2} H(t) dt \right)$, where ${\cal P}$ denotes path ordering.
\end{enumerate}

From these assumptions alone we cannot conclude that ${\cal A}_{t, \Delta t}$ contains the full set of observables, nor can we conclude that all
information is present at the boundary. However, given any observable ${\cal O}_{t_0} \in {\cal A}_{t_0, \Delta t}$, we can use (i) and (ii) to define a 1-parameter family of operators ${\cal O}_{t} \in {\cal A}_{t, \Delta t}$ which satisfy
\begin{equation}
\frac{d}{dt} {\cal O}_t = i[H(t), {\cal O}_t].
\end{equation}
It then follows from (ii) that $\frac{d}{dt} {\cal O}_t$ {\it also} lies in ${\cal A}_{t, \Delta t}$.  Since this holds for each possible ${\cal O}_{t} \in {\cal A}_{t, \Delta t}$, the algebra does not change with time.  I.e., each ${\cal A}_{t, \Delta t}$ contains the {\em same} set of observables.  In this sense, any information which happens to be present at the boundary at any time $t_1$ remains present at any
other time $t_2$. This result is naturally called  `boundary unitarity.'

We again stress that the above argument does not assume completeness of the boundary observables.  In particular, assumption (i) does not specify the Hilbert space on which $H(t)$ is self-adjoint.  We leave thus open the possibility of new non-perturbative bulk observables, or perhaps even of new observables corresponding to `baby universes.'  The role of assumption (i) is merely to ensure that the path-ordered exponential of $\int H(t) dt$ is well-defined.  In a corresponding argument at the classical level, all that would be required is that one be able to flow any boundary observable ${\cal O}$ by any finite amount of time along the (time-dependent) Hamiltonian vector field generated by $H(t)$; i.e., one simply requires time-evolution to be well-defined along the asymptotic boundary.  Such a requirement would amount to a rather weak form of cosmic censorship.

To provide some physical interpretation of the above result, consider a hypothetical observer who
lives outside the spacetime but who can interact with our spacetime
through the boundary observables.  If the observer has complete
control over the full algebra  ${\cal A}_{t, \Delta t}$ of boundary observables at each $t$, then at any time $t_2$ boundary unitarity will allow
her to extract any information which she has
encoded in the spacetime at any earlier time $t_1.$

Physically, the point is that particles which travel inward from the boundary at time $t_1$ leave an imprint on the boundary fields: the gravitational constraints precisely encode the total energy in the gravitational flux $\Phi$ at the boundary.  Because energy is the generator of time translations, the boundary observer can recover the desired information at any later time through appropriate couplings to this energy. Such processes will be explored in detail in \cite{soon}.

\section{Discussion}
\label{disc}

We have argued that perturbative quantum gravity about a collapsing
black hole background is, in a certain sense, holographic.  By this
we mean that, in the asymptotically flat context, the algebra
generated by asymptotic fields on $I^+$ within any neighborhood of
$i^0$ contains a complete set of observables. In the AdS context,
 the algebra of boundary observables associated with any
neighborhood of any Cauchy surface of the boundary spacetime is
similarly complete.  The fact that the gravitational Hamiltonian is
a pure boundary term played a key role, in a manner similar to that predicted in \cite{BMR}.

If this same algebra remains complete at the non-perturbative level, and  if perturbative quantum gravity about flat space is a
good approximation to some asymptotically flat non-perturbative
quantum gravity theory near past infinity ($i^-$ and $I^-$), future
infinity ($i^+$ and $I^+$), and spacelike infinity ($i^0$), it
follows  that the S-matrix is unitary.  This is again true if the completeness holds only in some appropriate superselection sector, as it would in an asymptotically flat analogue of the scenario outlined in \cite{BadH}.

 It is interesting to
classify possible failures of the assumption that perturbative gravity describes physics near $I^\pm$, $i^\pm$, and $i^0$  into two types.  First, the
physics might be described by perturbative quantum gravity about
some different background.  This might occur if the original
boundary conditions are somehow unstable and if additional
boundaries arise dynamically.  The other sort of failure would
preserve the boundary conditions but not allow a good approximation
by perturbative quantum gravity.  This might occur if, for example,
strongly coupled regions continue to interact with perturbative
fields at all times.  This could be the case in so-called
third-quantized theories \cite{3Q}, in which a given universe
continually interacts with a bath of baby universes.  However, in
such cases a form of unitarity may nevertheless hold due to the
superselection effects discussed in \cite{3QSS}.

In the AdS context, much weaker assumptions imply that similar
superselection effects {\em must} occur.  Specifically, whether or
not the set of boundary observables is complete, boundary unitarity
follows directly from the assumption that, in the non-perturbative
theory, the
algebra of boundary observables again contains a self-adjoint Hamiltonian.
While complete information may
never be present at the boundary, any information present there at one
time $t_1$ is also contained in boundary observables at any
other time $t_2$.  Any independent observables that may exist do not affect the evolution of boundary observables, though a
given quantum state might contain interesting correlations.  We note briefly that this fits well with the picture of certain extensions of AdS/CFT discussed e.g. in
\cite{Maldacena,FHMMRS} and with the general picture of AdS/CFT described in \cite{BadH}.

A number of possible objections were already addressed in the introduction.  Nonethelss, the reader may have certain further concerns.  For example, one may worry that the presence of so much information near
infinity might  violate the ``no quantum Xerox theorem'' \cite{noXerox}.   However, the original quantum state has in no way
been copied to new degrees of freedom.  Instead, the equations of
motion imply operator identities which require two a priori
different operators to be sensitive to the same qubit of quantum information.

One might also worry that our scenario may lead to paradoxes associated with
non-commuting measurements of some qubit being performed by
spacelike-separated observers: one in the interior of the spacetime
who measures local degrees of freedom, and one at the boundary who
makes use of the holographic encoding in the
algebra of boundary observables.  However, in a context where the boundary observables are complete,
the  the boundary observer has access to {\em all} degrees
of freedom, including the measuring devices of the local observer.
As a result, no paradoxes can arise. Any measurement made by a local
observer can always be undone by the boundary observer, though it would of course be interesting to understand the details.

An interesting, if perhaps somewhat artificial, context where the usual algebra of boundary observables is {\it not} complete can be constructed by adding a second boundary to the spacetime.  We may then place one observer outside each boundary, so that there is no danger of the local observer's devices being holographically encoded at the other boundary.  Since the interesting case arises when the two boundaries are in causal contact, we take this new boundary to be at finite distance (i.e., it is not an asymptotic boundary).

In the asymptotically flat version, this finite boundary may prohibit $i^+$ from being regular and may also interfere with the scattering of
wavepackets at early times.  As a result, we cannot conclude that
complete information is contained in a neighborhood of $i^0$.
However, at least in the AdS case our notion of boundary unitarity
will remain. Attempts to make use of this effect to extract a priori ``lost'' information appear to involve
extremely precise measurements of the gravitational flux $\Phi$ at
infinity. For now, we merely note that such experiments are very
difficult.  Indeed, we expect that the coarse-graining which leads
to semi-classical black hole thermodynamics is mostly a lack of
precision in measuring $\Phi$.  In this way, our perspective is
consistent with that of \cite{BMR}, and also with \cite{VB} (where information is also lost
simply by the erasure of quantum mechanical detail in semiclassical
measurements). This issue and the associated possible
paradoxes will be explored further in \cite{soon}.

There are many interesting issues that we have not addressed in this work.  For example, we have in no way suggested a microscopic mechanism that would determine
the entropy of black holes, or even to render it  finite.  As a result, we do not address the sort of unitarity
questions raised in \cite{Maldacena,Hawking}.

Even under the assumptions which led to unitarity of the S-matrix, a second (related) issue that we have not addressed is the {\em rate} at which information is transferred to the Hawking radiation.   To see the relation to the density of states, let us briefly summarize the picture of this process suggested by our arguments in the asymptotically flat context.  Motived by our perturbative results, we first assumed that the algebra of observables near $i^0$ is complete, and contains full information (at least in some superselection sector).  The most important observable was the gravitational flux $\Phi$, which led to completeness when combined with the usual perturbative observables.  However, an observer outside the black hole who uses, say, a set of particle detectors to extract information from the outgoing Hawking radiation does not measure $\Phi$ directly.  Instead, the flux of stress-energy in the Hawking radiation is related (via the gravitational Gauss' law) to the difference between $\Phi$ at $i^0$ and the corresponding gravitational flux $\Phi_{horizon}$ at the black hole horizon.   If one assumes that the density of states associated with $\Phi_{horizon}$ is given by the Bekenstein-Hawking formula, then one can predict the rate at which information is transferred to the Hawking radiation.  This amounts essentially to the classic analysis of \cite{Page}.  However, we again emphasize that we have provided no detailed justification for this assumption here.

What we have done is to point out that, if the black hole evaporates completely, the constraints then relate $\Phi$ directly to the stress tensor.  At this point there is no analogue of $\Phi_{horizon}$ and the information has become fully encoded in the Hawking radiation.   Furthermore, even before the black hole evaporates fully, we see that the horizon need not limit the transfer of information to outgoing radiation.  Since information associated with particle degrees of freedom inside the black hole is also encoded in the gravitational field outside the black hole (e.g., in $\Phi$), local physics outside the horizon is in principle sufficient to imprint this information on the Hawking radiation.

The essential point in our discussion was that the Hamiltonian of a
classical diffeomorphism-invariant theory is a pure boundary term.  A similar
feature holds in quantum perturbation theory, and it seems reasonable to
conjecture this property to hold in a non-perturbative quantum
theory -- even if the concepts of spacetime and
diffeomorphism-invariance themselves break down. This conjecture seems to hold, for example, in AdS/CFT  \cite{LargeN}, see \cite{HS,BK}.

As we have seen, the logical consequence of this property is that
the asymptotic fields store information in a way that would not be
possible in a local quantum field theory.  It is clear that such
arguments can be generalized to many other boundary conditions.
A generalization may also hold for the case of closed cosmologies.  There one imagines
that a physical clock might play the role of the boundaries used
above.  In perturbation theory, the gravitational constraints will
tie the energy of such a clock to the integral of the linearized
stress tensor of the gravitational degrees of freedom, so that it
might be used much like the gravitational flux $\Phi$ in our work
above.  Indeed, one might model such an observer by replacing their
worldline with an interior boundary.  We will save the detailed
exploration of such ideas for future work.

\appendix

\section{Tail Corrections in the Perturbative Holography Argument for Asymptotically Flat Spacetimes}
\label{app}

This appendix gives a more detailed discussion of how the "tail-terms" in the asymptotically flat perturbative holography argument of section \ref{pert}  are controlled for linear fields.  For interacting fields, it is clear that similar arguments hold at the level of perturbation theory.  We consider massless linear quantum field theory on a spherically symmetric asymptotically flat black hole background.  Below, we use the notation of massless scalar fields.  However, precisely the same equations hold for e.g. Maxwell or gravitational fields providing that one add a sufficient number of extra indices, that one replace scalar Green's functions with vector or tensor Green's functions, and that one replace the Klein-Gordon product with the appropriate vector or tensor symplectic product.  The field below will be called $h$, and can be viewed either as a scalar or as a tensor field with indices suppressed.

As in section \ref{pert}, we consider a field operator $h(v)$ on $I^-$.  We imagine that we have expanded the field in spherical harmonics, and that we consider a component with total angular momentum $l$.  (Thus, we might write $h(v)$ as $h^l(v)$, but we suppress the $l$ to simplify the notation.)
The field operator on $I^-$ is defined from the bulk field operator $h(x)$ by the rescaling necessary to make $h(v)$ finite on $I^-$.  It will simplify the notation to keep this rescaling implicit and to use to distinguish $h(x)$ from $h(v)$ only by the choice of argument.  I.e., any field evaluated on $I^\pm$ has been rescaled, while fields evaluated in the bulk (or on the black hole horizon) have not been rescaled.  We will use the same convention for the advanced/retarded Green's functions $G^\pm(x,y)$ (so that these Green's functions are also rescaled whenever one or both of their arguments lies on $I^\pm$, and where $G^\pm(x,y)$ also refer to angular momentum $l$).

Section \ref{pert} first translates $h(v)$ backward in time a distance $\tau$ along $I^-$.  We thus consider $h(v-\tau)$.  It then uses the equations of motion to express $h(v-\tau)$ in terms of fields on the future horizon (${\cal H}$) and on $I^+$.  As always for linear fields, this relation can be written in the form

\begin{equation}
\label{forward}
h(v - \tau) = \int_{x \in {\cal H} \cup I^+}  G^+(v-\tau, x)   (n^a  \stackrel
{\leftrightarrow} {\frac{ \partial}{\partial x^a} }) h(x) .
\end{equation}
where $n^a$ is an appropriately normalized (null) normal to the surfaces ${\cal H}, I^+$, and where the advanced Green's function is the relevant choice because $I^-$ is to the past of ${\cal H}, I^+$.  Here we choose sign conventions and such for the Green's functions to make the above statement true.

At this point, we wish to choose some (fixed) retarded time $u_1$ on $I^+$ and use it to divide $h(v-\tau)$ into two parts.  The part associated via (\ref{forward}) with $u < u_1$ on $I^+$ belongs to the algebra ${\cal A}_{u_1}^+$ of interest.  The rest is the error term from the tail that needs to be controlled.  We thus define the error term as

\begin{equation}
\label{forward}
\Delta(v - \tau, u_1) = \int_{x \in {\cal H}}  G^+(v-\tau, x)   (n^a  \stackrel
{\leftrightarrow} {\frac{ \partial}{\partial x^a} }) h(x) +  \int_{x \in {I^+}, u(x) \ge u_1}  G^+(v-\tau, x)   (n^a  \stackrel
{\leftrightarrow} {\frac{ \partial}{\partial x^a} }) h(x) .
\end{equation}

To evaluate the contributions of $\Delta(v-\tau, u_1)$, it is useful to {\rm again} express $\Delta(v-\tau, u_1)$ in terms of the original fields on $I^-$ by again using the free field Green's functions\footnote{The reason that this is useful is that, since we specify the state on $I^-$, it is easier to understand the effect of 
$\Delta(v-\tau, u_1)$ when expressed in terms of operators on $I^-$.}.  We have

\begin{eqnarray}
\label{forback}
\Delta(v - \tau, u_1) &=& \int_{x \in {\cal H}} \int_{y \in I^-} G^+(v-\tau, x)   (n^a  \stackrel
{\leftrightarrow} {\frac{ \partial}{\partial x^a} }) G^-(x, y)   (n^a  \stackrel
{\leftrightarrow} {\frac{ \partial}{\partial y^a} })  h(y) \cr &+&  \int_{x \in {I^+}, u(x) \ge u_1}  \int_{y \in I^-} G^+(v-\tau, x)   (n^a  \stackrel
{\leftrightarrow} {\frac{ \partial}{\partial x^a} }) G^-(x, y)   (n^a  \stackrel
{\leftrightarrow} {\frac{ \partial}{\partial y^a} })  h(y) .
\end{eqnarray}

To package this result in a transparent way, note that (\ref{forback}) may be written in the form

\begin{equation}
\Delta(v - \tau, u_1) =\int_{y \in I^-}  \Delta(v - \tau, u_1, y) (n^a  \stackrel
{\leftrightarrow} {\frac{ \partial}{\partial y^a} })  h(y)  ,
\end{equation}
where we have defined 

\begin{equation}
\label{Dcoeffs}
\Delta(v - \tau, u_1, y )=\int_{x \in {\cal H}}  G^+(v-\tau, x)   (n^a  \stackrel
{\leftrightarrow} {\frac{ \partial}{\partial x^a} }) G^-(x, y) +  \int_{x \in {I^+}, u(x) \ge u_1} G^+(v-\tau, x)   (n^a  \stackrel
{\leftrightarrow} {\frac{ \partial}{\partial x^a} }) G^-(x, y)  .
\end{equation}

Our goal is to show that the coefficients $\Delta(v - \tau, u_1, y )$ vanish as some power law in $v-\tau$ in the limit $\tau \rightarrow \infty$ for each $y$.  A simple way to do so is to note that since both ${\cal H}$ and $I^+$ are null surfaces, the null normals $n^a$ are in fact tangent to the respective surfaces.  Thus, we may perform integrations by parts so that all of the derivatives act on factors of $G^-(x, y)$.  I.e, we may write

\begin{eqnarray}
\label{Dcoeffs2}
\Delta(v - \tau, u_1, y )&=& 2 \int_{x \in {\cal H}}  G^+(v-\tau, x)   (n^a  
 {\frac{ \partial}{\partial x^a} }) G^-(x, y) +  \int_{x \in {I^+}, u(x) \ge u_1} G^+(v-\tau, x)   (n^a  
 {\frac{ \partial}{\partial x^a} }) G^-(x, y) \cr  &+& {\rm boundary \ term \ at \ {\cal H} \ vertex },
\end{eqnarray}
where (due to Price's law \cite{GPP,DR}) the only non-zero boundary term is at the ``vertex'' of the future horizon (i.e., the point at which, due to spherical symmetry, all horizon generators meet in a caustic).

We now note that, due to Price's law \cite{GPP,DR}, the function $(n^a  
 {\frac{ \partial}{\partial x^a} }) G^-(x, y)$ falls off at least as $u^{-2}$ as the retarded time $u = u(x)$ becomes large along $I^+$,  and that it falls off at least as $v^{-2}$ as the retarded time $v = v(x)$ becomes large along ${\cal H}$.  Thus, even if the  $G^+(v-\tau, x)$ factor were not present, the above integrals would converge absolutely at large $u(x), v(x)$.

It is now easy to see that $\Delta(v - \tau, u_1, y )$ vanishes in the limit $\tau \rightarrow \infty$ for each $y$.  One simply notes that, the Green's function $ G^+(v-\tau, x) $ vanishes as a power law for large positive $\tau$ (in particular, for $v - \tau \ll v_1(u_1, L)$ as defined by figure \ref{translate} at each fixed $x$ on either ${\cal H}$ or $I^+$.  This is just the statement that, if there is a source in the distant past of fixed strength, the amount of radiation it emits which reaches any fixed point on either ${\cal H}$ or $I^+$ vanishes in the limit that the source acts only at very large times before the black hole forms.

Since integrals of the form $\int dz f(z) g(z)$ with $f(z)$ bounded and $g(z) \in L^1$ vanish in any limit where $f(z) \rightarrow 0$ pointwise, it follows that $\Delta(v - \tau, u_1, y )$ vanishes in the limit $\tau \rightarrow \infty$ for each $y$ as desired.    Thus the error operator  $\Delta(v - \tau, u_1)$ of (\ref{forback}) vanishes weakly in this limit and, as argued in section \ref{pert}, so does 
$e^{-i \tau \Phi} \Delta (v-\tau,u_1)e^{i \tau \Phi}$.

The above analysis concerned free (linear) fields.  Section \ref{pert} is also interested in interacting fields, but only at the perturbative level.  At each order in perturbation theory one may proceed along lines similar to the above, but with the number of Green's functions increasing at each order in perturbation theory.  It therefore clear that perturbative corrections will behave similarly to the linearized results above; i.e., the error term will vanish as desired in the limit $\tau \rightarrow \infty$.

\subsection*{Acknowledgements}
The author has benefited from discussions with numerous physicists
at UCSB, the Perimeter Institute, the ICMS workshop on Gravitational
Thermodynamics and the Quantum Nature of Space Time in Edinburgh,
and the ICTS Monsoon Workshop on String theory in Mumbai.  In
particular, he is grateful to Alejandra Castro, Sergei Gukov, Steve
Giddings, Gary Horowitz, Veronika Hubeny, Joe Polchinski, Mukund
Rangamani, Rafael Sorkin, and Lenny Susskind, Aron Wall, and
especially to Ted Jacobson and Simon Ross for their
thought-provoking comments and questions and to Maulik Parikh for a
careful reading of an earlier draft. This work was supported in part
by the US National Science Foundation under Grant No.~PHY05-55669,
and by funds from the University of California. The author thanks
the Tata Institute for Fundamental Research and the International
Center for Theoretical Sciences for their hospitality and support
during the final stages of this project.

\end{document}